\documentclass[12pt]{article}
\usepackage{amsmath}
\usepackage{graphicx,psfrag,epsf}
\usepackage{enumerate}
\usepackage{natbib}
\usepackage{hyperref}
\usepackage{rotating}
\usepackage{pdflscape}

\newcommand{\blind}{1}

\usepackage{color}
\usepackage{mathtools}
\usepackage{bm}
\usepackage{array}
\usepackage{amssymb}
\bibpunct{(}{)}{;}{a}{}{,}

\begin{document}

	\def\spacingset#1{\renewcommand{\baselinestretch}%
		{#1}\small\normalsize} \spacingset{1}

	
	\if1\blind
	{
		\title{\bf Inference for Hit Enrichment Curves, with Applications to Drug Discovery}
		\author{Jeremy R Ash\textsuperscript{1,2} \thanks{
				JRA gratefully acknowledges support from NIH training grant T32ES007329, the Triangle Center for Evolutionary Medicine, and SAS Institute.}
			\ \ and 
			Jacqueline M. Hughes-Oliver\textsuperscript{1}\\
			\textsuperscript{1} Bioinformatics Research Center, Department of Statistics,\\ North Carolina State University, Raleigh, NC\\
			\textsuperscript{2} JMP Division, SAS Institute, Cary, NC}

		\maketitle
	} \fi
	
	\if0\blind
	{
		\bigskip
		\bigskip
		\bigskip
		\begin{center}
			{\LARGE\bf Inference for Hit Enrichment Curves, with Applications to Drug Discovery}
		\end{center}
		\medskip
	} \fi
	
	\bigskip
	\begin{abstract}
		In virtual screening for drug discovery, hit enrichment curves are widely used to assess the performance of ranking algorithms with regard to their ability to identify early enrichment. Unfortunately, researchers almost never consider the uncertainty associated with estimating such curves before declaring differences between performance of competing algorithms. Appropriate inference is complicated by two sources of correlation that are often overlooked: correlation across different testing fractions within a single algorithm, and correlation between competing algorithms. Additionally, researchers are often interested in making comparisons along the entire curve, not only at a few testing fractions. We develop inferential procedures to address both the needs of those interested in a few testing fractions, as well as those interested in the entire curve. For the former, four hypothesis testing and (pointwise) confidence intervals are investigated, and a newly developed EmProc approach is found to be most effective. For inference along entire curves, EmProc-based confidence bands are recommended for simultaneous coverage and minimal width. Our inferential procedures trivially extend to enrichment factors, as well.
	\end{abstract}
	
	\noindent%
	{\it Keywords:}  Virtual Screening; Enrichment Factor; Lift Curve; Early Enrichment; Ranking Algorithm; Empirical Process
	\vfill
	
	\thispagestyle{empty}
	\setcounter{page}{0}
	
	\newpage
	\spacingset{1.45} 
\section{Introduction}
\label{sec:intro}

Ranking algorithms order items according to the belief that they possess some desired feature. When the presence/absence of the desired feature is known, ranking algorithms are often evaluated by ``testing'' items according to the relative rank or testing order. Ideally, all early tests reveal the desired feature. The statistics and machine learning communities use a number of performance curve variants to evaluate ranking algorithms, including the hit enrichment curve and the enrichment factor or lift curve. Popular software such as the R package \verb+caret+ \citep{McCollum2009}, SAS Enterprise Miner \citep{SAS}, and JMP \citep{SAS} can be used to construct these curves. Some curves are used extensively in the evaluation of virtual screens of chemical compounds for drug discovery \citep{Geppert2010}. They are also used in a number of other applications such as the evaluation of marketing campaigns \citep{Rosset2001}.

In the context of virtual screening, the desired feature is often a biological activity. Typically the desired activity is binding to a protein target, so we will refer to the chemical compounds as ligands. Ligands are scored, where the scores are provided by one of the many ranking algorithms available, such as molecular docking algorithms, pharmacophore models, or
quantitative structure-activity relationship (QSAR) models. 

\cite{Empereur-Mot2016} recently provided a web application for constructing performance curves to evaluate virtual screening methods. One of the performance curves they provide is the hit enrichment curve. The software provides many nice interactive features for exploring performance metrics at points along a curve. It also provides utilities for combining the scores from multiple methods into a consensus score that may improve ranking performance. 

Of tremendous importance in a world where virtual screening data is tightly guarded, \cite{Empereur-Mot2016} make some of their data freely available. We use one of the case studies as a demonstration dataset. The target was the protein regulating gene peroxisome proliferator-activated receptor gamma (PPARg), which has been linked to several diseases such as obesity, diabetes, atherosclerosis, and cancer \citep{PPARG}. While this is a small dataset for a retropsective screening evaluation by present-day standards, the rarity of actives ($\hat{\pi}_+=0.0265$ is the observed fraction of active ligands) and the small testing fractions are very consistent with larger screening studies \citep{Zhu2013}. The primary goal in such a study is to discover scoring methods that are able to correctly identify active ligands, and to do so very early in the testing phase \citep{Truchon2007, Jain2008, Geppert2010}.

The hit enrichment curve is commonly used to summarize effectiveness of a screening campaign. It plots the proportion of active ligands identified (i.e., the recall) as a function of the fraction of ligands tested, where testing order is based on the score produced by a virtual screening method. Larger recall values are preferred and, going a step further, these are more relevant when they occur at small testing fractions. In other words, one hopes to demonstrate improvement in \textit{early enrichment}.

While the hit enrichment curve is designed to show results for all testing fractions, it is common to focus on fractions below 0.1 and even below 0.001 \citep{Zhu2013}. The size of our demonstration dataset limits us to consider 0.001 as the smallest testing fraction (resulting in just three ligands tested), but present-day screening campaigns can easily involve millions of ligands and hence testing fractions below 0.001 are reasonably considered. For a given testing fraction, one may want to compare observed recall values for competing scoring methods.

For the PPARg study, \cite{Empereur-Mot2016} use three popular docking methods to score ligands: Surflex-dock, ICM, and Vina. We scale all docking scores so that a larger value is consistent with active ligands; as a result, ICM and Vina scores have been negated. \cite{Empereur-Mot2016} also evaluate several methods for constructing consensus scores, and we limit investigations to two of their best performers: maximum of the z-scores from Surflex-dock and ICM, and the minimum of the ranks from Surflex-dock and ICM.

\begin{figure}[h]
	\centering\includegraphics[width=.8\textwidth]{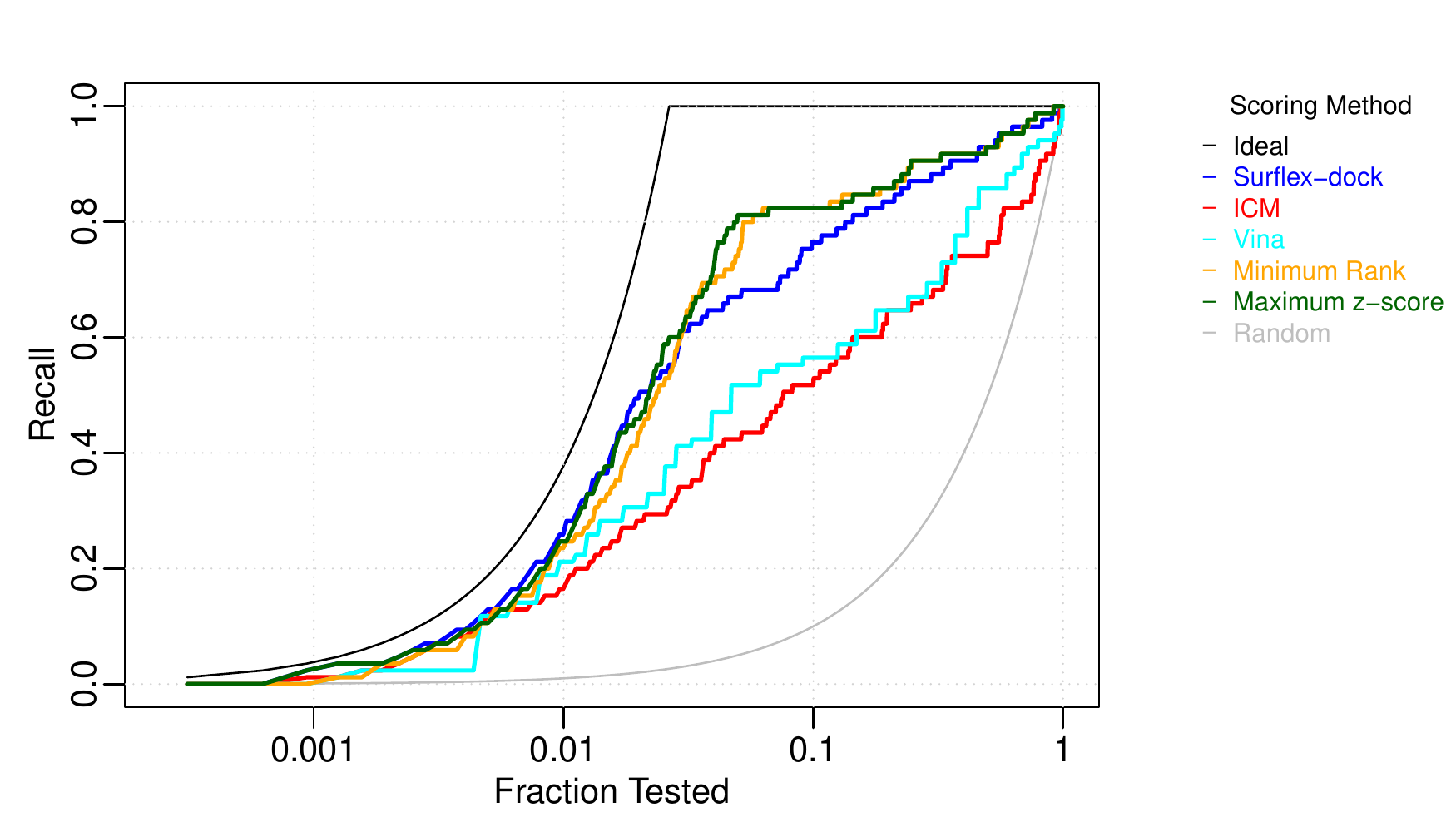}
	\caption{Hit enrichment curves for the PPARg application, comparing five scoring methods over the full range of testing fractions. Ideal and random hit enrichment curves are also shown.\label{fig:5recalls}}
\end{figure}

Figure \ref{fig:5recalls} shows estimated hit enrichment curves for the five scoring methods over the full range of testing fractions, along with the ideal hit enrichment curve that would be expected if the 85 active ligands were identified in the first 85 tests, and the hit enrichment curve consistent with random identification of active ligands. This figure differs slightly from that produced by the web application of \cite{Empereur-Mot2016} due to a difference in how ties are handled; we use inverse distribution functions (strategy discussed below) to avoid arbitrary indications of better performance due simply to random ordering. 

Comparing consensus methods to individual docking methods, is the observed improvement in recall at testing fraction 0.1 significant? With recall being the proportion of active ligands identified at this testing fraction of 321 tests, it may be tempting to apply a procedure based on comparing independent binomial proportions. There are, however, two complications. The first is that determination of testing order requires scores from all ligands, and hence this introduces correlation between the 321 tests that are applied for a single scoring method. The second complication is that competing scoring methods are very likely positively correlated and so the uncertainty associated with differences between hit enrichment curves may possibly be reduced, thus improving the power to uncover differences. This paper develops appropriate techniques for comparing hit enrichment curves that result from competing scoring methods. Our inferential procedures trivially extend to enrichment factors, as well. The need for proper inferential methods for these early enrichment metrics has been emphasized by many recent papers \citep{nicholls2014confidence,robinson2020validating}.

Section \ref{sec:ranking} introduces hit enrichment curves constructed from ranking algorithms.
Section \ref{sec:compare2} presents four approaches for hypothesis testing and confidence intervals to compare two hit enrichment curves, along with simulation studies to compare effectiveness of the approaches; EmProc is newly proposed here, while three other approaches are applied in new ways.
Section \ref{sec:bands1} presents confidence band procedures and simulation results for an entire hit enrichment curve from a single algorithm, while Section \ref{sec:bands2} considers bands for the difference between two hit enrichment curves.
Section \ref{sec:pparg} revisits the PPARg application.
Section \ref{sec:discuss} includes a discussion of general findings from this study, and also makes connections to the broader task of assessment of virtual screening campaigns.

\section{Evaluation of Ranking Algorithms} \label{sec:ranking}

Let $S$ denote the score from a ranking algorithm, where larger values of $S$ suggest stronger belief that a ligand is active. $S$ is reasonably regarded as a random variable. Activity of a ligand may also be regarded as a random variable: $X = I(active)$, where $I(\cdot)$ is the indicator function. That is, $X=1$ when a ligand belongs to the active class ($+$) and $X = 0$ when a ligand belongs to the inactive class ($-$). Let $P(X = 1) = \pi_+$. Given that a ligand is active, $S$ has cumulative distribution function $F_+(s)$, and given that a ligand is not active, $S$ has cumulative distribution function $F_{-}(s)$. Combining the scores from both classes results in the mixture distribution
$	F_S(s) = \pi_+F_+(s) + (1-\pi_+)F_-(s). $

Once ligands have been ranked according to their score, $S$, a threshold on the score, $t$, will prioritize a top fraction of the data set for testing. Let this top fraction be $r = P(S > t)$, which is the x-axis of the population hit enrichment curve. The hit enrichment curve (also known as the enrichment curve, accumulation curve, or percent captured response curve) is often used when an entire curve is used to evaluate a virtual screening campaign. Population hit enrichment curves plot $P(S > t | +)$ on the y-axis, where $P(S>t|+)$ is known as $recall$ at threshold $t$.

So far, we have described the population level distributions of random variables $X$ and $S$, that is, the distribution of ligand activity for the population of drug candidates from which a data set is sampled and the distribution of scores that a ranking algorithm would assign to them. We consider a data set under examination to be a random sample of activity and score pairs $\{(X_i,S_{i}); i = 1, ..., n\}$ from these population level distributions. Let $\{S_{i}^+; i = 1, ..., n^+\}$ be the scores that were sampled from the $+$ class mixture component, $F_+(s)$, and $\{S_{i}^-; i = 1, ..., n^-\}$ be the scores that were sampled from the $-$ class mixture component, $F_-(s)$.

The empirical hit enrichment curve plots the cumulative fraction of actives on the y-axis, identified as a function of the top $r$ fraction of ranked ligands. This means all compounds with scores beyond the percentile $100(1-r)$ are ``tested" and the cumulative fraction of actives determined. Another way of determining this percentile would be to choose a threshold $\hat{t}_r$ such that the fraction of items with $S > \hat{t}_r$ is $r$. We use $\hat{t}$ instead of $t$ to denote that this threshold defines a fraction of the sample data and not the population.

Specifically, we define, $\hat{F}(\cdot)$ and $\hat{F}_+(\cdot)$ to be the empirical cumulative distribution functions (cdfs) for all and $+$ scores:
\begin{align*}
	\hat{F}(s) &= \frac{1}{n}\sum_{i=1}^{n}I(S_i \leq s), &
	\hat{F}_+(s) &= \frac{1}{n^+}\sum_{i=1}^{n^+}I(S_i^+ \leq s).
\end{align*}
Given a testing fraction $r$, we ideally choose a threshold $\hat{t}_r$ to be the score percentile selected for testing such that $r = 1-\hat{F}(\hat{t}_r)$. To accommodate the possible existence of ties in the observed data on scores, we  define $\hat{t}_r=\min\{t:\hat{F}(t)\geq 1-r\}$. Estimated recall at testing fraction $r$ is the fraction of the active compounds that are correctly predicted to be active (i.e. $S_{i}^+ > \hat{t}_r$) and is thus obtained as $\widehat{\theta}_r = 1-\hat{F}_+(\hat{t}_r) = 1-\hat{F}_+(\hat{F}^{-1}(1-r))$. Thus, the empirical hit enrichment curve plots the pairs $\{r, \widehat{\theta}_r\}$. 

\cite{Jiang2015} have shown that if $\pi_+ \in (0, 1)$, then the empirical hit enrichment curve is an unbiased estimator of the population hit enrichment curve. 

\section{Compare Hit Enrichment Curves from Competing Algorithms}\label{sec:compare2}

We wish to determine whether one ranking algorithm has significantly better performance than another at a given testing fraction $r$. For $r=1-F(t_r)$, let $\theta_r=P(S>t_r|+)$ denote the true population-level recall for a ranking algorithm at testing fraction $r$. Let $\{(X_i, S_{1i}, S_{2i}); i = 1, ..., n\}$ be a random sample from the ligand activity distribution and the score distributions of ranking algorithm 1 and 2. We assume that the triplets $(X_i, S_{1i}, S_{2i})$ are independent across $i$. However, $S_{1i}$ and $S_{2i}$ are likely correlated. The amount of correlation will depend on the extent to which the scores are measures of the same characteristics of the ligands. For example, competing docking scoring functions are often parameterized in similar ways (e.g., Glide SP and Glide XP; see \cite{Friesner2004}), and competing QSAR models often utilize highly correlated sets of descriptors. Using the random sample, we estimate the difference in performance between two algorithms at a given $r$, $\hat{\theta}_{1r} - \hat{\theta}_{2r}$, and perform a hypothesis test to determine if the difference is significant.

First considering a single algorithm, let $Q_r=\sum_{i=1}^n X_iI(S_i>\hat{t}_r)$ represent the number of active ligands that are examined at testing fraction $r$. We estimate recall at testing fraction $r$ using $\hat{\theta}_{r}=Q_r/\sum_{i=1}^n X_i$, noting that both the numerator and denominator are random variables. The activity rate $\pi_+$ is reasonably estimated as $\hat{\pi}_+=\sum_{i=1}^n X_i/n$, so an alternative expression for estimated recall is $\hat{\theta}_{r}=Q_r/(n\hat{\pi}_+)$. Because $\hat{t}_r$ is estimated using the entire dataset through the empirical cdf $\hat{F}(\cdot)$, $Q_r$ is not binomially distributed. To properly account for this, \cite{Jiang2015} take an empirical process approach to derive asymptotic normality of $\hat{\theta}_{r}$. Their result is that $\sqrt{n}(\widehat{\theta}_r - \theta_r) \xrightarrow[]{d} N(0, \tau^2_{\theta_r})$ for $r \in (0, 1)$ as $n \rightarrow \infty$, with corresponding asymptotic variance expression
\begin{equation}
	\tau^2_{\theta_r}/n \coloneqq Var_{JZ}(\widehat{\theta}_r) = Var_B(\widehat{\theta}_r)\Big[1 - 2\Lambda_r + \frac{\Lambda^2_r(1-r)r}{{\pi}_+\theta_r(1-\theta_r)}\Big]\label{eq:VarJZ}, 
\end{equation}
where 
\begin{equation}
	Var_B(\widehat{\theta}_r)=(n{\pi}_+)^{-1}\theta_r(1-\theta_r) \label{eq:VarB}
\end{equation}
is the simple binomial variance, and $\Lambda_r = P(+|S = t_r)$ is a threshold-specific activity rate. This result assumes that $\pi_+ > 0$, and that the conditional densities $f_+(s)$ and $f_-(s)$ are positive and continuously differentiable in a neighborhood of $S=t_r$; henceforth called Conditions 1 and 2.

When it comes to comparing estimated recall across two competing algorithms, there are two sources of correlation that should be addressed. One source is correlation induced by needing to estimate $t_r$ using $\hat{t}_r$, and that is addressed by using the result of \cite{Jiang2015}. The other source of correlation arises because competing algorithm scores are derived using some common data, and this source of correlation has not been previously addressed in the literature. By accounting for both types of correlation, we expect to improve the power to detect real differences in algorithmic performance.

The following subsections present four methods of testing for significant differences in recall for two competing algorithms. First, we extend the approach of \cite{Jiang2015} to use an empirical process approach that accounts for the correlation \textit{between} algorithms, in addition to the correlation induced \textit{within} a particular algorithm by estimation of $t_r$; this method is called EmProc. Second, we present details on how a McNemar procedure for correlated proportions may be applied to the application; as far as we know, this has not been previously done. Third, we apply the \cite{Jiang2015} result for hypothesis testing; this method is called IndJZ and is only optimized to address correlation within each algorithm but not correlation between algorithms. And fourth, we treat $Q_{1r}$ and $Q_{2r}$ as correlated binomial random variables, thus ignoring correlations induced by estimating $t_r$ but accounting for correlations between algorithms; this method is called CorrBinom. The four methods are compared using a simulation study in Section \ref{sec:simu}.

All four methods are based on asymptotic normality of a test statistic of the form
\begin{align*}
	Z_r = \frac{\widehat{\theta}_{1r} - \widehat{\theta}_{2r}}{SE(\widehat{\theta}_{1r} - \widehat{\theta}_{2r})}.
\end{align*}
We reject $H_{0r}: \theta_{1r} = \theta_{2r}$ if $|Z_r| > z_{\alpha/2}$, where $z_{\alpha/2} = 1.96$ for an $\alpha = .05$ level test.  Pointwise confidence intervals are obtained as 
\begin{equation}
	(\widehat{\theta}_{1r} - \widehat{\theta}_{2r}) \pm z_{\alpha/2}SE(\widehat{\theta}_{1r} - \widehat{\theta}_{2r}). \label{eq:pointCI}
\end{equation}
The methods differ in their approach to estimating $Var(\widehat{\theta}_{1r} - \widehat{\theta}_{2r})$.

\subsection{EmProc: Adjust for correlation between algorithms and correlation within each algorithm}\label{sec:EmProc}

Taking an empirical process approach, the functional delta method employed by \cite{Jiang2015} was extended to derive the following asymptotic normality result concerning $\widehat{\theta}_{1r} -\widehat{\theta}_{2r}$.

\noindent\textbf{Theorem 1:} Given that Conditions 1 and 2 are satisfied for both $\theta_{1r}$ and $\theta_{2r}$, then
\[ \sqrt{n}\Bigr\{(\widehat{\theta}_{1r} -\widehat{\theta}_{2r}) - (\theta_{1r} - \theta_{2r})\Bigl\} \xrightarrow[]{d} N(0, \tau^2_{\theta_{1r}, \theta_{2r}}) \]
as $n \rightarrow \infty$. Furthermore, the asymptotic variance expression is 
\begin{equation} 
	\tau^2_{\theta_{1r}, \theta_{2r}}/n \coloneqq Var_{\mbox{\scriptsize EmProc}}(\widehat{\theta}_{1r} -\widehat{\theta}_{2r}) = Var_{JZ}(\widehat{\theta}_{1r}) + Var_{JZ}(\widehat{\theta}_{2r}) - 2Cov_{\mbox{\scriptsize EmProc}}(\widehat{\theta}_{1r} ,\widehat{\theta}_{2r}), \label{eg:VarEmProc}
\end{equation}
where:  $Var_{JZ}(\cdot)$ is as given in equation (\ref{eq:VarJZ}) and applied for each algorithm;
\begin{align} 
	Cov_{\mbox{\scriptsize EmProc}}(\widehat{\theta}_{1r} ,\widehat{\theta}_{2r}) 
	&= Cov_{B}(\widehat{\theta}_{1r} ,\widehat{\theta}_{2r})\left\{(1-\Lambda_{1r}-\Lambda_{2r}) + \frac{(\gamma_{12\cdot r} - r^2)\Lambda_{1r}\Lambda_{2r}}{\pi_+(\theta_{12\cdot r} - \theta_{1r}\theta_{2r})}\right\}; \nonumber\\
	Cov_{B}(\widehat{\theta}_{1r} ,\widehat{\theta}_{2r}) 
	&= (n{\pi}_+)^{-1}\Bigl( \theta_{12\cdot r} - \theta_{1r}\theta_{2r} \Bigr) \label{eq:CovB}
\end{align}
is the covariance between binomial counts; $r=P(S_j>t_{jr})$ and determines the threshold for algorithm $j$, for $j=1,2$; $\theta_{12\cdot r} = P(S_1>t_{1r},S_2>t_{2r}|+)$ is the conditional probability that both algorithms result in testing an active ligand because it is highly ranked by both algorithms; $\gamma_{12\cdot r} = P(S_1>t_{1r},S_2>t_{2r})$ is the unconditional probability that a ligand is highly ranked by both algorithms; and $\Lambda_{jr} = P(+|S_j = t_{jr})$ for $j=1,2$. Details and derivations are in the Appendix. 

To estimate $Var_{\mbox{\scriptsize EmProc}}(\widehat{\theta}_{1r} -\widehat{\theta}_{2r})$, we replace the population parameters with consistent estimators. The frequency distribution for the tested/not-tested status according to both ranking algorithms for the active ligands is shown in Table \ref{tab:mcnemar}, where 
$Q_{jr} = \sum_{i=1}^{n}X_iI\{S_{ji} > \widehat{t}_{jr} \}$ counts the number of active ligands tested by algorithm $j$ for $j=1,2$, and
$Q_{12\cdot r} = \sum_{i=1}^{n}X_iI\{S_{1i} > \hat{t}_{1r} , S_{2i} > \hat{t}_{2r} \}$
counts the number of active ligands tested by both algorithms. As previously discussed, we estimate $\theta_{jr}$ with $\widehat{\theta}_{jr}=Q_{jr}/(n\widehat{\pi}_+)$. Additional estimates are obtained as $\widehat{\gamma}_{12\cdot r} = \sum_{i=1}^n I\{S_{1i} > \hat{t}_{1r} , S_{2i} > \hat{t}_{2r} \}/ n$, $\widehat{\theta}_{12\cdot r} = Q_{12\cdot r}/(n\widehat{\pi}_+)$, and $\widehat{\Lambda}_{jr}$ is obtained using Nadaraya-Watson kernel regression  with the ``rule-of-thumb" bandwidth selector \citep{li2007nonparametric}.

\begin{table}
  \caption{Frequency distribution for the tested/not-tested status of active ligands according to both ranking algorithms, at a fixed testing fraction $r$.\label{tab:mcnemar}}
  \begin{center}
	\begin{tabular}{c|ccc}
		& Algorithm 2 tested  & Algorithm 2 not-tested  & Row total \\
		\hline
		Algorithm 1 tested & $Q_{12\cdot r}$ & $Q_{1r} - Q_{12\cdot r}$ & $Q_{1r}$\rule{0pt}{2.7ex}\\
		Algorithm 1 not-tested & $Q_{2r} - Q_{12\cdot r}$  & $n\hat{\pi}_+ - (Q_{1r} + Q_{2r} - Q_{12\cdot r})$ & $n\hat{\pi}_+ - Q_{1r}$ \\
		Column total & $Q_{2r}$ & $n\hat{\pi}_+ - Q_{2r}$ & $n\hat{\pi}_+$ \\
	\end{tabular}
  \end{center}
\end{table}

Under the null hypothesis $H_{0r}: \theta_{1r}=\theta_{2r}$, we could alternatively use a pooled estimator of $\theta_{jr}$ for $j \in \{ 1,2\}$, namely
$\widehat{\theta}_r = \frac{1}{2}(\widehat{\theta}_{1r} + \widehat{\theta}_{2r})$ to replace both $\widehat{\theta}_{1r}$ and $\widehat{\theta}_{2r}$ in variance expression (\ref{eg:VarEmProc}). We consider variance estimates using both the unpooled and pooled approaches.

\subsection{McNemar's test for difference in recall}\label{sec:mcnemar}

When estimating recall, the same set of active ligands simultaneously serves as the set of ``trials" for both ranking algorithms, with the decision being whether or not each algorithm selects the active ligand for testing. The consequence is that the data for both algorithms may be viewed as fully paired. The standard test used for paired proportions is McNemar's test \citep{Agresti2007, Fagerland2013}.  Table~\ref{tab:mcnemar} shows how the number of active ligands
tested by either ranking algorithm can be written as a $2 \times 2$ contingency table. The estimated recall values present themselves as the marginal probabilities of testing an active ligand for each ranking algorithm. Consequently, the asymptotic McNemar test is based on test statistic 
\[ Z_r = \frac{((Q_{1r} - Q_{12\cdot r}) - (Q_{2r} - Q_{12\cdot r}))}{\sqrt{(Q_{1r} - Q_{12\cdot r}) + (Q_{2r} - Q_{12\cdot r})}} = \frac{(Q_{1r} - Q_{2r})}{\sqrt{Q_{1r} + Q_{2r} - 2Q_{12\cdot r}}}
= \frac{(\widehat{\theta}_{1r} - \widehat{\theta}_{2r})}{\sqrt{Q_{1r} + Q_{2r} - 2Q_{12\cdot r}} / (n\widehat{\pi}_+)} \]
and it assumes that discordant counts $(Q_{2r} - Q_{12\cdot r})$ and $(Q_{1r} - Q_{12\cdot r})$ are large. In a  simulation study comparing the asymptotic McNemar test to a number of other tests for paired nominal data, the asymptotic McNemar test was found to be the most powerful across simulation scenarios, though slightly liberal in terms of type I error \citep{Fagerland2013}.

While the asymptotic McNemar test enforces the null condition $\theta_{1r}=\theta_{2r}$ to replace $Q_{1r} - Q_{2r}$ with zero in the variance expression, an alternative approach is needed to obtain pointwise confidence intervals. Pointwise Wald confidence intervals use the following standard error expression in equation (\ref{eq:pointCI}):
\[ SE(\widehat{\theta}_{1r} - \widehat{\theta}_{2r}) = \sqrt{Q_{1r} + Q_{2r} - 2Q_{12\cdot r} -(Q_{1r} - Q_{2r})^2/(n\widehat{\pi}_+) }/(n\widehat{\pi}_+). \]
Unfortunately, several studies \citep{Newcombe1998,Fagerland2013,RodriguezdeGil2013} demonstrate inadequate coverage properties of the Wald interval. The Bonett-Price \citep{BonettPrice2012} adjusted interval is a simple modification of the Wald interval, and it has been shown to have good coverage properties \citep{Fagerland2013,RodriguezdeGil2013}. We refer to the Bonett-Price adjustment as a ``plus" adjustment because it adds one unit to each of the discordant counts shown in Table~\ref{tab:mcnemar}, then applies the Wald formula. More precisely, discordant count $Q_{1r} - Q_{12\cdot r}$ becomes $Q_{1r} - Q_{12\cdot r}+1$ and discordant count $Q_{2r} - Q_{12\cdot r}$ becomes $Q_{2r} - Q_{12\cdot r}+1$, thus adding one to each of the marginal counts and two to the overall total. As a result, the Bonett-Price plus interval is 
\begin{eqnarray*}
	\frac{ Q_{1r} - Q_{2r} }{n\widehat{\pi}_+ {+2}} \pm z_{\alpha/2}
	\sqrt{ \frac{1}{(n\widehat{\pi}_+ {+2})^2} \left[ 
		\left( Q_{1r} + Q_{2r} - 2Q_{12\cdot r} {+2} \right)  -
		\frac{ \left(Q_{1r} - Q_{2r}\right)^2 }{(n\widehat{\pi}_+ {+2})} 
		\right] } \label{eq:BPci} ,
\end{eqnarray*}
noting that both the center point and the standard error have been adjusted.

\subsection{IndJZ: Adjust for correlation within each algorithm but not correlation between algorithms}\label{sec:IndJZ}

If we assume that $\widehat{\theta}_{1r}$ and $\widehat{\theta}_{2r}$ are independent, then
$Var_{JZ}(\widehat{\theta}_{1r} - \widehat{\theta}_{2r}) = Var_{JZ}(\widehat{\theta}_{1r}) + Var_{JZ}(\widehat{\theta}_{2r})$, where $Var_{JZ}(\widehat{\theta}_{jr})$ is obtained as in equation (\ref{eq:VarJZ}) for $j=1,2$. 
For testing equality of recall for the algorithms, either a pooled or unpooled estimator of the variance could be used, as previously discussed.

When the competing algorithms have scores that are highly positively correlated, it is expected that the IndJZ approach will lead to standard errors that are unnecessarily large, resulting in an underpowered test.

\subsection{CorrBinom: Adjust for correlation between algorithms but not correlation within each algorithm}\label{sec:CorrBinom}

In this approach, we treat $Q_{jr}$ as if it follows a simple binomial distribution, even though it does not. As a result, the relevant variance expression is 
$ Var_B(\widehat{\theta}_{1r} -\widehat{\theta}_{2r}) = Var_{B}(\widehat{\theta}_{1r}) + Var_{B}(\widehat{\theta}_{2r}) - 2Cov_{B}(\widehat{\theta}_{1r} ,\widehat{\theta}_{2r}), $
where $Var_B(\cdot)$ is defined in equation (\ref{eq:VarB}) and $Cov_B(\cdot,\cdot)$ is defined in equation (\ref{eq:CovB}).

For testing equality of recall for the algorithms, either a pooled or unpooled estimator of the variance could be used, as previously discussed.

\subsection{Simulation Results\label{sec:simu}}

\subsubsection{Simulation of Benchmarking Data Sets\label{sec:bench}}

\cite{Geppert2010} and \cite{Xia2015} have recently reviewed the standard data sets used to benchmark virtual screening tools. The goal in designing a benchmark data set is to mimic real world chemical collections -- this means that the score and activity distributions of the compounds in the benchmark should resemble these populations. 


Prospective virtual screens are typically conducted on large databases such as ZINC \citep{Sterling2015}. These databases can be considered random samples from ``drug like'' chemical space. The performance curves estimated by retrospective virtual screens will likely mis-estimate performance in a prospective virtual screen if the benchmark data set is not also a random sample. To this date, directory of useful decoys (DUD) is the most widely used collection of benchmarking data sets used in the evaluation of retrospective virtual screens, however each of the biases mentioned have been observed in these data sets \citep{Irwin2008,Good2008,Xia2015}. The directory of useful decoys, enhanced (DUD-E) data sets \citep{Mysinger2012} were developed to address some biases in these data sets, though the data sets still lack the experimental testing of decoys (i.e., there is still false negative bias) and there is an unrealistic frequency of actives included in each of the data sets. The MUV benchmarking data sets \citep{Rohrer2009} have also been developed with the intention of minimizing these biases. A clear advantage over DUD-E is that the decoys in MUV have been tested experimentally. The authors of MUV collected 18 primary high-throughput screen assays from PCBioAssay \citep{NCBIResourceCoordinators2016}. Actives were further confirmed with low throughput assays to minimize the number of false positives, and additional checks for false negatives were performed. We modeled our simulations on the MUV benchmarking data sets, because we believe these data sets to be the most representative of the population of drug candidates.

Basing our simulations on MUV, we simulated data sets with $n = 150,000$, $\pi_+ = .002$, and $skew = 499$ (where $skew$ is ratio of inactives to actives, $(1-\pi_+)/\pi_+$). In general, this is representative of a typical virtual screening data set with large sample sizes and extreme class imbalance.

\subsubsection{Study Design\label{sec:simuDesign}}

A study was conducted to investigate the power of EmProc, McNemar, IndJZ, and CorrBinom. For each of EmProc, IndJZ, and CorrBinom, a total of four modifications were considered according to whether pooling was applied or whether the Bonett-Price plus adjustment was applied. Although the Bonett-Price plus adjustment was proposed for the Wald confidence interval to improve coverage probability, we wondered if it might also be used for hypothesis testing. McNemar was not modified. In fact, McNemar is identical to CorrBinom with pooling and without plus adjustment.

A study was also conducted to investigate coverage probabilities of the confidence intervals associated with EmProc, IndJZ, CorrBinom, and the Bonett-Price plus-adjusted Wald interval (the latter being referred to as the McNemar inverval, for brevity). There is no justification for pooling when considering confidence intervals, so no pooling is applied. We continue to investigate the impact of the Bonett-Price plus adjustment. The McNemar interval is identical to the CorrBinom interval with plus adjustment.

Four data-generating mechanisms were considered: binormal or bibeta; and correlation of 0.9 or 0.1. In the binormal model, both algorithms have $F_-$ as the standard normal distribution function, while Algorithm 1 has $F_+$ as the normal with mean $0.8\sqrt{2}$ and variance one, and Algorithm 2 has $F_+$ as the normal with mean $0.6\sqrt{2}$ and variance one. Algorithm 1 has a relatively large separation (Cohen's $D=0.8$) between the score distributions of the $+$ and $-$ classes; Algorithm 2 has a slightly diminished performance (Cohen's $D=0.6$). To capture the fact that both algorithms are scoring the same compounds, we simulated the positive scores from a bivariate normal with marginals as described above and correlation parameter $\rho=0.1$ or $\rho=0.9$. The negative scores were simulated from a separate bivariate normal with the described marginals and the same correlation parameter. In the bibeta model, both algorithms have $F_-$ as the beta distribution with $\alpha=2$ and $\beta=5$, while Algorithm 1 has $F_+$ as the beta(5,2) and Algorithm 2 has $F_+$ as the beta(4,2). Sampling was done using the bivariate beta distributions to incorporate correlations similar to the binormal model. Sampling was conducted using the \verb+copula+ R package \citep{copula2020}.

There is much greater separation between $F_-$ and $F_+$ in the bibeta model, so the true hit enrichment curves are higher than in the binormal model, and we expect greater correlation of scores across early testing fractions. But for both the binormal and bibeta models, parameters were chosen to result in very similar hit enrichment curves for the two competing algorithms, thus creating a challenging task for hypothesis testing of differences between hit enrichment curves.

Studies were conducted using 10,000 Monte Carlo replicates. We estimated the type I error rate for the hypothesis test methods assuming that both ranking algorithms had either the score distributions of Algorithm 1 or Algorithm 2.

\subsubsection{Results\label{sec:simuResults}}

\underline{\em Hypothesis Testing}

Pooling has no impact on either the power or protection from type I errors for EmProc, IndJZ, or CorrBinom (results not shown for brevity), so we limit further discussion to the versions of these tests constructed without pooling. The impact of the Bonett-Price plus adjustment on power is mixed (results not shown). The plus adjustment had no noticeable impact on power under the bibeta model. But under the binormal model, the plus adjustment caused a noticable decrease in power for a medium range of tests performed (approximately 30 to 500). For this reason we limit further discussion to the versions of these tests constructed without plus adjustment. As a reminder, McNemar is based on pooling and no plus adjustment.

\begin{figure}
	\centering\includegraphics[width=.8\textwidth]{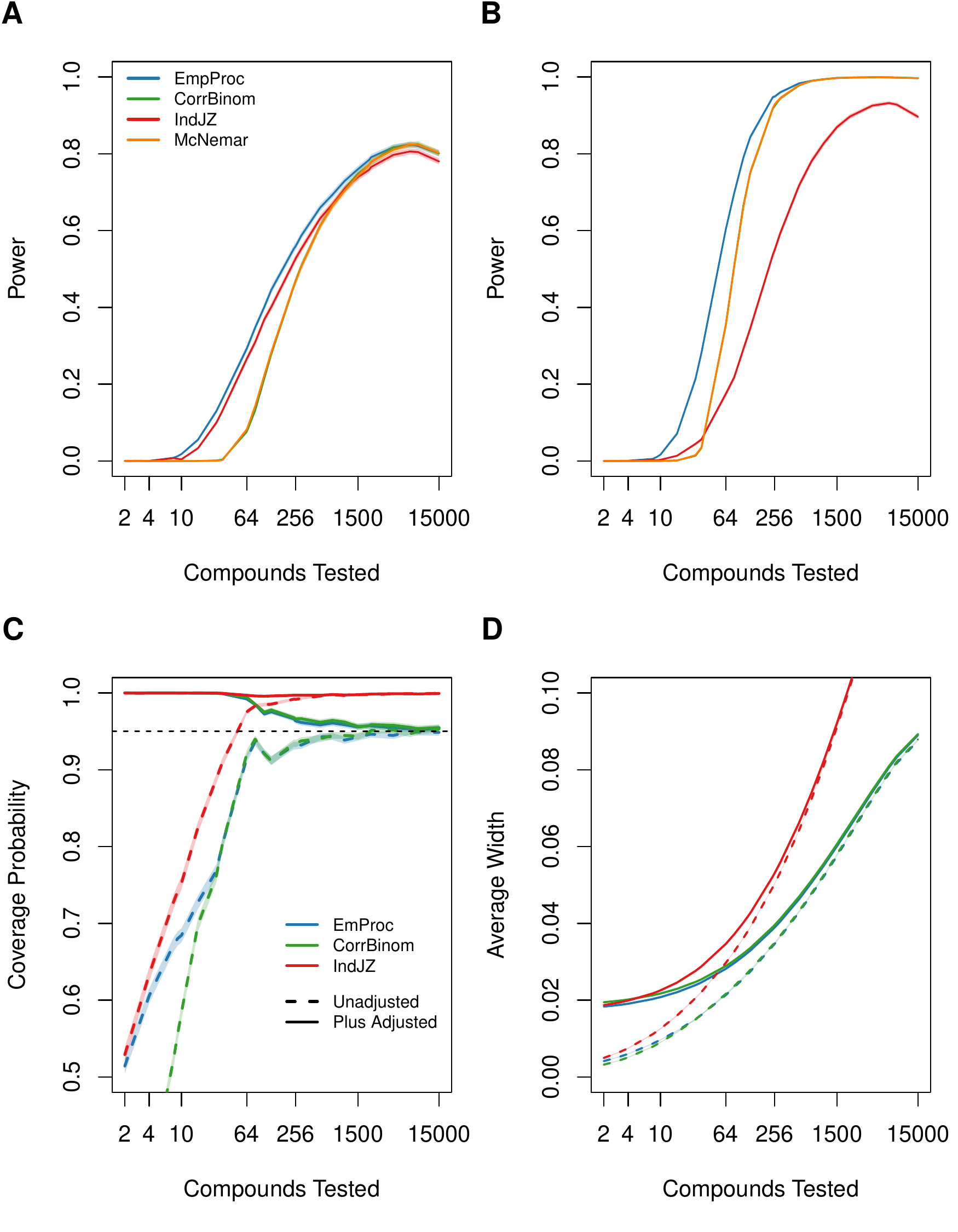}
	\caption{Comparison of EmProc, CorrBinom, IndJZ, and McNemar in terms of hypothesis testing and pointwise confidence intervals to compare hit enrichment curves for competing algorithms. (A-B): Estimated power of the hypothesis test to detect differences between two competing algorithms, where each algorithm follows a bibeta model and scores are correlated with $\rho=.1$ (A) or $\rho=.9$ (B). (C-D): Estimated coverage probability (C) and average width (D) of pointwise confidence intervals for the difference in hit enrichment curves for two competing algorithms, where each algorithm follows a binormal model and scores are correlated with $\rho=.9$. Simulations were conducted with 10,000 Monte Carlo replicates. Shading show the Monte Carlo estimate $\pm$ 1.96 times the Monte Carlo standard error.\label{fig:simuHTCI}}
\end{figure}

Figure \ref{fig:simuHTCI} shows estimated power curves for correlation of 0.1 (A) and 0.9 (B) under the bibeta model. We focus on results for number of tests ($nr$) ranging from two tests to testing ten percent of the total sample size; in practice, screening campaigns will interrogate only a tiny fraction of the available virtual screening library \citep{Zhu2013}. The CorrBinom and McNemar approaches are noticeably suboptimal. In the presence of weak correlation, EmProc and IndJZ have comparable performance. EmProc dominates in the presence of strong correlation between scores. Indeed, EmProc is the only approach that is designed to address both the correlation between scores of competing algorithms, and the correlation that is induced within a particular algorithm as a result of having to estimate thresholds. The binormal model (results not shown) yielded similar findings. Type I error rates (results not shown) are well controlled to their nominal values of 0.05.

For hypothesis testing, we recommend EmProc without pooling and without plus adjustment, because EmProc has the greatest power compared to IndJZ, CorrBinom, and McNemar, while maintaining control of type I error rates. If one chooses to use either IndJZ or CorrBinom, the unpooled and non-plus-adjusted versions should be used.

\underline{\em Confidence Intervals}

Figure \ref{fig:simuHTCI} shows estimated coverage probabilities (C) and average widths (D) for confidence intervals EmProc, IndJZ, and CorrBinom, based on the binormal model with correlation of 0.9. The McNemar interval is equivalent to the CorrBinom interval with plus adjustment, so while the figure does not explicitly include the label of McNemar, it is included.

The most obvious finding is that the Bonett-Price plus adjustment dramatically improves coverage probabilities when the number of tests is small; this is because the plus adjustment results in wider intervals when the number of tests is small. As the number of tests increase, the plus and no-plus versions converge, and they approach nominal coverage. 
By not accounting for correlation across competing algorithms, IndJZ standard errors are unnecessarily large, resulting in wide intervals that provide conservative coverage. The plus-adjusted versions of EmProc and CorrBinom (and hence also McNemar) provide conservative coverage when the number of tests is small, but coverage approaches the nominal level as number of tests increase.

For confidence intervals, we recommend the plus-adjusted version of EmProc, because it is best able to balance achieving nominal coverage rates while minimizing the width of confidence intervals. And if other procedures are used, the plus adjustment should be used as well.

\section{Confidence Bands}\label{sec:bands}

\subsection{Bands for a Single Algorithm}\label{sec:bands1}

\subsubsection{Methods}\label{sec:bands1method}

An ideal scenario would be to accompany hit enrichment curves, such as those shown in Figure 1, with confidence regions. Non-overlapping regions would provide an alternative justification for claiming significant differences between competing algorithms. Let $\bm{\theta}$ denote the vector of recall values from a single algorithm at the vector $\bm{r}=(r_1,r_2,\ldots,r_k)$ of $k$ ordered testing fractions $r_1<r_2<\cdots<r_k$. We seek a $100(1-\alpha)$ percent confidence region for $\bm{\theta}$. While the pointwise confidence interval approach of equation (\ref{eq:pointCI}) could be modified using a Bonferroni adjustment, such corrections are known to be conservative when $k$ is large, leading to unnecessarily wide intervals.

In their technical report, \cite{Jiang2014} suggested an alternate confidence band estimation procedure, and gave brief comments on simulation results, but some details were omitted. We complete these details to state the following result. Under the previously mentioned Conditions 1 and 2, as $n\rightarrow\infty$, 
$ \sqrt{n}( \widehat{\bm{\theta}}-\bm{\theta}) \stackrel{d}{\longrightarrow} N(\bm{0},\bm{V}), $
where $\widehat{\bm{\theta}}=\left( \widehat{\theta}_{r_1},\ldots,\widehat{\theta}_{r_k} \right)$ is the vector of recall estimators as previously defined, and $\bm{V}=\{ V_{ij} \}_{i,j=1,\ldots,k}$. Moreover,
$ V_{ii}/n = Var_{JZ}(\widehat{\theta}_{r_i}), $
and, for $r_i<r_j$,
\begin{equation}
	V_{ij}/n = Cov_{\mbox{\scriptsize EmProc}}(\widehat{\theta}_{r_i} ,\widehat{\theta}_{r_j}) =  \frac{\theta_{r_i}\left( 1-\theta_{r_j} \right)}{n{\pi}_+} \left\{(1-\Lambda_{r_i}-\Lambda_{r_j}) + \frac{r_i(1-r_j)\Lambda_{r_i}\Lambda_{r_j}}{\pi_+ \theta_{r_i}\left( 1-\theta_{r_j} \right) }\right\}.
	\label{eq:AN2}
\end{equation}
Derivation details are omitted because they are similar to the steps in the appendix; see \cite{ash2020methods} for further details. To provide a working distribution for $\widehat{\bm{\theta}}$, an estimator of $\bm{V}$ is obtained by replacing population parameters with consistent estimators. This working distribution is the basis of our approximate confidence regions.

Our most straight-forward approach is to use a Wald 100($1-\alpha$) percent confidence ellipsoid, defined as
$
	\left\{ {\bm{\mathbf{\MakeLowercase{\theta}}}} \ : \ n({\bm{\mathbf{\MakeLowercase{\theta}}}}  - \widehat{{\bm{\mathbf{\MakeLowercase{\theta}}}}})^T \widehat{{\bm{\mathbf{\MakeUppercase{V}}}}}^{-1} ({\bm{\mathbf{\MakeLowercase{\theta}}}}  - \widehat{{\bm{\mathbf{\MakeLowercase{\theta}}}}}) \leq \chi^2_{k, 1-\alpha} \right\},
$
where $\chi^2_{k, 1-\alpha}$ is the $1-\alpha$ percentile
of the chi-squared distribution with $k$ degrees of freedom. But the Wald confidence ellipsoid does not produce regions that are of the rectanguloid form
$
	\left\{ {\bm{\mathbf{\MakeLowercase{\theta}}}} \ : \ \widehat{\theta}_i \pm q\cdot SE(\widehat{\theta}_i) \ \forall \ i \in \{1, ..., k \} \right\}.
$
We have chosen to use confidence regions with a rectangular structure (i.e., a confidence band) and not ellipsoids because this allows confidence regions of high dimensions to be easily visualized.

Clearly, Bonferroni regions are rectanguloid, with $q=\sqrt{\chi^2_{1, 1-\alpha/k}}$. We mention two additional rectanguloid regions, following the naming conventions in \cite{MontielOlea2019}: the $\theta$-projection and sup-t bands. $\theta$-projection bands are obtained by identifying the smallest rectanguloid that contains the Wald ellipsoid, and results in $q=\sqrt{\chi^2_{k, 1-\alpha}}$. Upon further inspection, it becomes clear that $\theta$-projection bands are always at least as wide as Bonferroni bands, so they are not considered further. On the other hand, sup-t bands are the smallest rectanguloid that maintains the simultaneous coverage probability of $1-\alpha$, and are expected to have good performance. Their critical value $q$ must be obtained using Monte Carlo sampling. Briefly,
\begin{eqnarray*}
	1-\alpha & \leq & \Pr\left( |\widehat{\theta}_i - \theta_i| \leq q \cdot SE(\widehat{\theta}_i) \ \forall \ i \in \{1, \ldots, k \} \right) 
	= \Pr\left( \sup_{i=1, \ldots, k}\frac{|\widehat{\theta}_i - \theta_i|}{SE(\widehat{\theta}_i)} \leq q \right).
\end{eqnarray*}
Monte Carlo sampling is used to estimate $q$ as the $(1-\alpha)100$ percentile for the distribution of $\sup_{i=1, \ldots, k}|\widehat{\theta}_i - \theta_i| / SE(\widehat{\theta}_i)$.

\subsubsection{Simulation Results}

A study compared coverage probabilities achieved by confidence bands constructed using sup-t and Bonferroni approaches. Results are shown for both the standard (formulas shown in Section \ref{sec:bands1method}) and plus-adjusted versions of sup-t and Bonferroni bands. The familiar trick of ``add two successes and add two failures" \citep{Agresti1998}  before estimating proportions is what we refer to as the plus adjustment for bands corresponding to a single hit enrichment curve, not the Bonett-Price plus adjustment for comparing two algorithms that was used earlier. 
We computed bands for $\bm{\theta}$ using a grid of number of compounds tested between two and 15,000. We considered 25 points on the grid, defined as: $2^k$ for $k=1,\ldots,13$; $3^k$ for $k=1,\ldots,8$;  105, 300, 1500, 15000. 

\begin{figure}
	\centering\includegraphics[width=.8\textwidth]{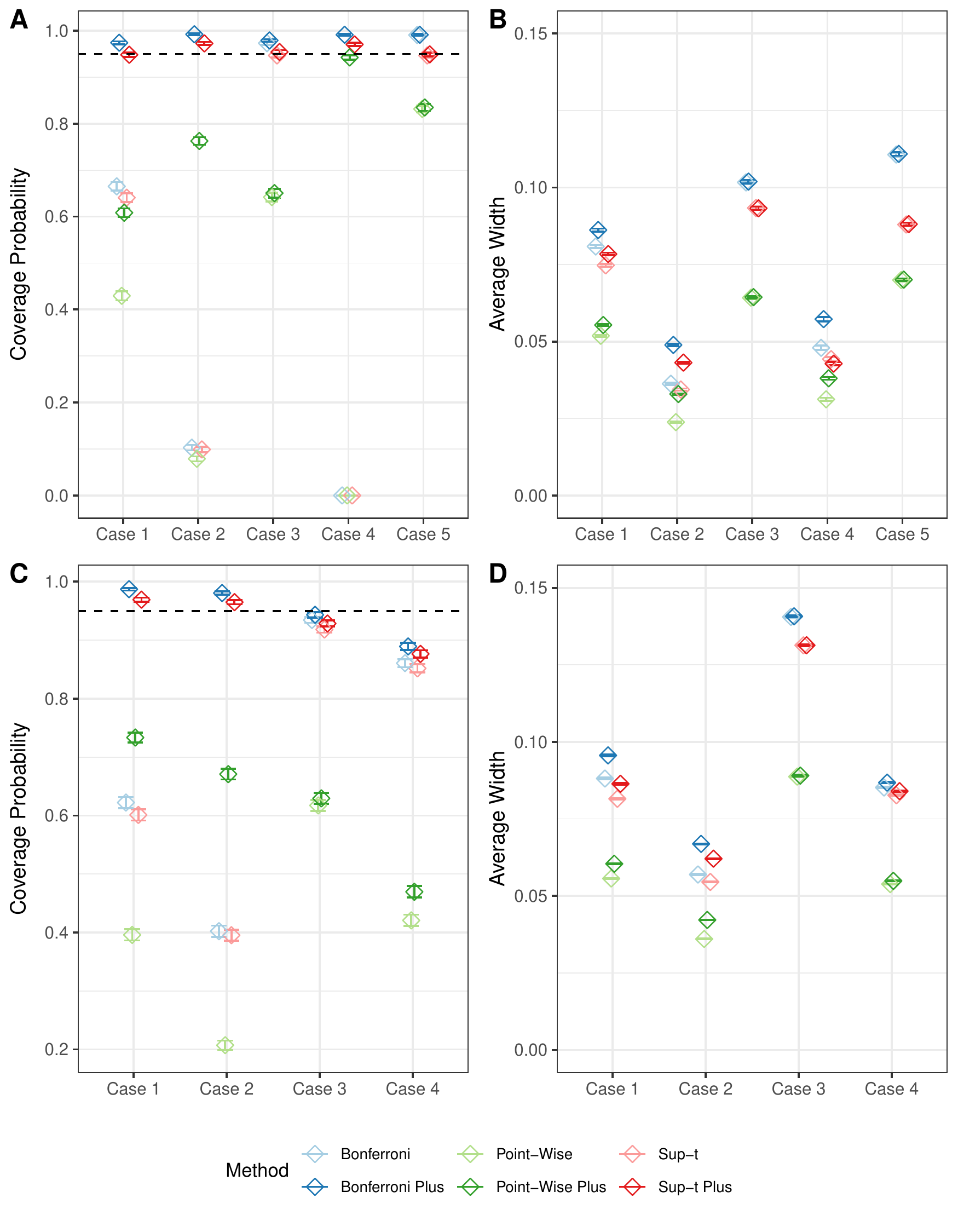}
	\caption{Comparison of sup-t and Bonferroni confidence bands. (A-B): Estimated coverage probability (A) and average width (B) of confidence bands for hit enrichment curves for a single algorithm, where the algorithm is generated from five different cases. (C-D): Estimated coverage probability (C) and average width (D) of confidence bands for the difference between two hit enrichment curves generated under four scenarios.  Simulations were conducted with 10,000 Monte Carlo replicates. Error bars show the Monte Carlo estimate $\pm$ 1.96 times the Monte Carlo standard error.\label{fig:simuBands}}
\end{figure}

Figure \ref{fig:simuBands} shows estimated coverage probabilities (A) and average widths (B) of confidence bands created based on five distributional cases. The distributional cases represent varying degrees of separation between $+$ and $-$ classes, and are chosen to mimic real-world scenarios. Case 1 is a binormal model, with equal unit variance and means zero and 1.4. Case 2 is another binormal model, with equal unit variance and means zero and 0.5; Case 2 offers much less separation than Case 1, so Case 2 results in lower values of recall. Case 3 is a bibeta model, with Beta(2,5) and Beta(5,2) distributions. Case 4 is another, more separated, bibeta model, with Beta(1,20) and Beta(20,1) distributions. Case 5 is made up of distributions of limited extent, namely uniform on (0,0.75) and uniform on (0.25,1).

The plus-adjusted Bonferroni bands have the highest coverage, but they are also the widest. The plus-adjusted sup-t bands are not as wide, yet have excellent coverage. As such, for confidence bands applied to a single hit enrichment curve, we recommend the plus-adjusted sup-t bands.

\subsection{Bands for the Difference Between Competing Algorithms}\label{sec:bands2}

\subsubsection{Methods}

While the pointwise confidence intervals of Section \ref{sec:compare2} offer effective comparisons of competing algorithms at a few selected testing fractions, it may be more desirable to perform comparisons across a large range of testing fractions. This may be accomplished by converting the pointwise confidence intervals into confidence bands, in much the same way that confidence bands were obtained in Section \ref{sec:bands1}. 

Letting $\bm{\theta}_\ell$ denote the vector of recall values from Algorithm $\ell$ ($\ell=1,2$), the method is based on the asymptotic result  
$ \sqrt{n} \left( (\widehat{\bm{\theta}}_1 - \widehat{\bm{\theta}}_2) - (\bm{\theta}_1 - \bm{\theta}_2) \right) \stackrel{d}{\longrightarrow} N(\bm{0},\bm{V}), $
where $n\rightarrow\infty$, and for $r_i\le r_j$,
\begin{equation}
	V_{ij}/n = 
	Cov_{\mbox{\scriptsize EmProc}}(\widehat{\theta}_{1r_i} ,\widehat{\theta}_{1r_j}) +
	Cov_{\mbox{\scriptsize EmProc}}(\widehat{\theta}_{2r_i} ,\widehat{\theta}_{2r_j}) -
	Cov_{\mbox{\scriptsize EmProc}}(\widehat{\theta}_{1r_i} ,\widehat{\theta}_{2r_j}) -
	Cov_{\mbox{\scriptsize EmProc}}(\widehat{\theta}_{1r_j} ,\widehat{\theta}_{2r_i}) .
	\label{eq:AN3}
\end{equation}
The first two components of equation (\ref{eq:AN3}) are obtained using equation (\ref{eq:AN2}), and the latter two components are obtained using 
\begin{equation}
	Cov_{\mbox{\scriptsize EmProc}}(\widehat{\theta}_{1r_i} ,\widehat{\theta}_{2r_j}) =  \frac{\left( \theta_{12\cdot r_i r_j} -\theta_{1r_i}\theta_{2r_j} \right)}{n{\pi}_+} \left\{(1-\Lambda_{1r_i}-\Lambda_{2r_j}) + \frac{\left( \gamma_{12\cdot r_i r_j} -r_i r_j \right) \Lambda_{1r_i}\Lambda_{2r_j}}{\pi_+ \left( \theta_{12\cdot r_i r_j} -\theta_{1r_i}\theta_{2r_j} \right) }\right\},
	\label{eq:AN4}
\end{equation}
where $\theta_{12\cdot r_i r_j}=P(S_1>t_{1r_i},S_2>t_{2r_j}|+)$ and $\gamma_{12\cdot r_i r_j}=P(S_1>t_{1r_i},S_2>t_{2r_j})$ are the conditional and unconditional probabilities that both algorithms test a ligand because it is highly ranked by both algorithms, albeit at different testing fractions $r_i$ and $r_j$. Equation (\ref{eq:AN4}) does not impose any restrictions between testing fractions $r_i$ and $r_j$.

As described in Section \ref{sec:bands1}, matrix $\bm{V}$ is estimated and used to construct sup-t and Bonferroni bands.

\subsubsection{Simulation Results}

Similar to Section \ref{sec:compare2}, a study was conducted to compare coverage probabilities and average widths of confidence bands constructed using sup-t and Bonferroni approaches under four settings of two competing algorithms: binormal or bibeta, and $\rho=0.1$ or 0.9. Bands were computed for $\bm{\theta}$ using a grid of size 25, with number of tested compounds being: $2^k$ for $k=1,\ldots,13$; $3^k$ for $k=1,\ldots,8$;  105, 300, 1500, 15000.

Figure \ref{fig:simuBands} shows estimated coverage probabilities (C) and average widths (D). Results are very similar to those observed for confidence bands for a single algorithm, namely that the plus-adjusted sup-t bands provide the best balance between coverage probabilities and average width.

For confidence bands applied to the difference between two hit enrichment curves, we recommend the plus-adjusted sup-t bands for achieving nominal coverage rates and minimizing width. The covariance used in constructing these bands arise from the EmProc approach.

\section{Revisit the PPARg Application}\label{sec:pparg}

For testing fractions 0.001, 0.01, and 0.1, Table \ref{tab:applic} provides details for all pairwise comparisons between scoring methods Surflex-dock (the best docking method), ICM (the worst docking method), and the maximum z-score (the best consensus method). For each of the three pairs, we provide the estimated difference between hit enrichment curves. Standard errors and resulting $p$-values are provided for the EmProc approach to conducting inference, and also for the remaining approaches McNemar, IndJZ, and CorrBinom. 

Acknowledging the multiple-testing scenario required to compare two scoring methods, Table \ref{tab:applic} also provides multiplicity-adjusted $p$-values. Given choice of a particular approach to inference (for example, EmProc), there are nine tests based on three pairs of scoring methods and three testing fractions. A simple Benjamini-Hochberg \citep{Benjamini1995} step-up procedure is used to control the false discovery rate. At testing fraction 0.1 (321 tests), the difference between Surflex-dock and the consensus method is not significant, but ICM is significantly worse than both Surflex-dock and the consensus method. These conclusions are clearly supported by all inferential approaches. At testing fraction 0.01 (32 tests), no significant differences are observed, but the EmProc procedure is seen to produce the smallest standard errors. With our relatively small dataset, testing fraction 0.001 results in only three tests, and there is too much uncertainty to make a reliable conclusion.

\begin{landscape}
\begin{table}
	\caption{Pairwise comparison of scoring methods surflex-dock (surf), ICM, and their consensus (maxz), at three testing fractions decided \textit{a priori}. For each pair of scoring methods, differences and standard errors of estimated hit enrichment curves are shown, along with raw and Benjamini-Hochberg adjusted $p$-values.\label{tab:applic}}
	\begin{center}\footnotesize
		\begin{tabular}[t]{lccc>{}c|ccc>{}c|cccc}
			\multicolumn{1}{c}{\em{\textbf{ }}} & \multicolumn{4}{c}{\em{\textbf{N tested = 3}} $(r=0.001)$} & \multicolumn{4}{c}{\em{\textbf{N tested = 32}} $(r=0.01)$} & \multicolumn{4}{c}{\em{\textbf{N tested = 321}} $(r=0.1)$} \\
			\cline{2-5} \cline{6-9} \cline{10-13}
			& Diff & Std Err & Raw p & Adj p & Diff & Std Err & Raw p & Adj p & Diff & Std Err & Raw p & Adj p\\\hline
			\multicolumn{13}{l}{\textbf{EmProc}}\\
			\hspace{1em}maxz - surf & 0.0000 & 0.0005 & 1.000 & 1.000 & -0.0118 & 0.0237 & 0.6200 & 0.6970 & 0.0588 & 0.0254 & 2.07e-02 & 6.21e-02\\
			\hspace{1em}maxz - icm & 0.0118 & 0.0143 & 0.410 & 0.527 & 0.0824 & 0.0402 & 0.0407 & 0.0733 & 0.3060 & 0.0541 & 1.60e-08 & 1.44e-07\\
			\hspace{1em}surf - icm & 0.0118 & 0.0142 & 0.409 & 0.527 & 0.0941 & 0.0429 & 0.0281 & 0.0632 & 0.2470 & 0.0626 & 7.91e-05 & 3.56e-04\\
			\multicolumn{13}{l}{\textbf{McNemar}}\\
			\hspace{1em}maxz - surf & 0.0000 & 0.0000 & 1.000 & 1.000 & -0.0118 & 0.0311 & 0.705 & 0.794 & 0.0588 & 0.0255 & 2.53e-02 & 7.60e-02\\
			\hspace{1em}maxz - icm & 0.0118 & 0.0203 & 0.564 & 0.725 & 0.0824 & 0.0557 & 0.144 & 0.260 & 0.3060 & 0.0552 & 2.07e-06 & 1.86e-05\\
			\hspace{1em}surf - icm & 0.0118 & 0.0203 & 0.564 & 0.725 & 0.0941 & 0.0614 & 0.131 & 0.260 & 0.2470 & 0.0642 & 3.86e-04 & 1.74e-03\\
			\multicolumn{13}{l}{\textbf{IndJZ}}\\
			\hspace{1em}maxz - surf & 0.0000 & 0.0138 & 1.000 & 1.000 & -0.0118 & 0.0497 & 0.8130 & 0.915 & 0.0588 & 0.0609 & 3.34e-01 & 5.28e-01\\
			\hspace{1em}maxz - icm & 0.0118 & 0.0143 & 0.411 & 0.528 & 0.0824 & 0.0482 & 0.0874 & 0.197 & 0.3060 & 0.0668 & 4.74e-06 & 4.26e-05\\
			\hspace{1em}surf - icm & 0.0118 & 0.0143 & 0.409 & 0.528 & 0.0941 & 0.0471 & 0.0458 & 0.137 & 0.2470 & 0.0693 & 3.63e-04 & 1.64e-03\\
			\multicolumn{13}{l}{\textbf{CorrBinom}}\\
			\hspace{1em}maxz - surf & 0.0000 & 0.0000 & 1.000 & 1.000 & -0.0118 & 0.0311 & 0.705 & 0.793 & 0.0588 & 0.0255 & 2.12e-02 & 6.35e-02\\
			\hspace{1em}maxz - icm & 0.0118 & 0.0203 & 0.563 & 0.724 & 0.0824 & 0.0557 & 0.139 & 0.251 & 0.3060 & 0.0552 & 3.07e-08 & 2.76e-07\\
			\hspace{1em}surf - icm & 0.0118 & 0.0203 & 0.563 & 0.724 & 0.0941 & 0.0614 & 0.125 & 0.251 & 0.2470 & 0.0642 & 1.20e-04 & 5.40e-04\\
	\end{tabular}
	\end{center}
\end{table}
\end{landscape}

\begin{figure}
	\centering\includegraphics[width=\textwidth]{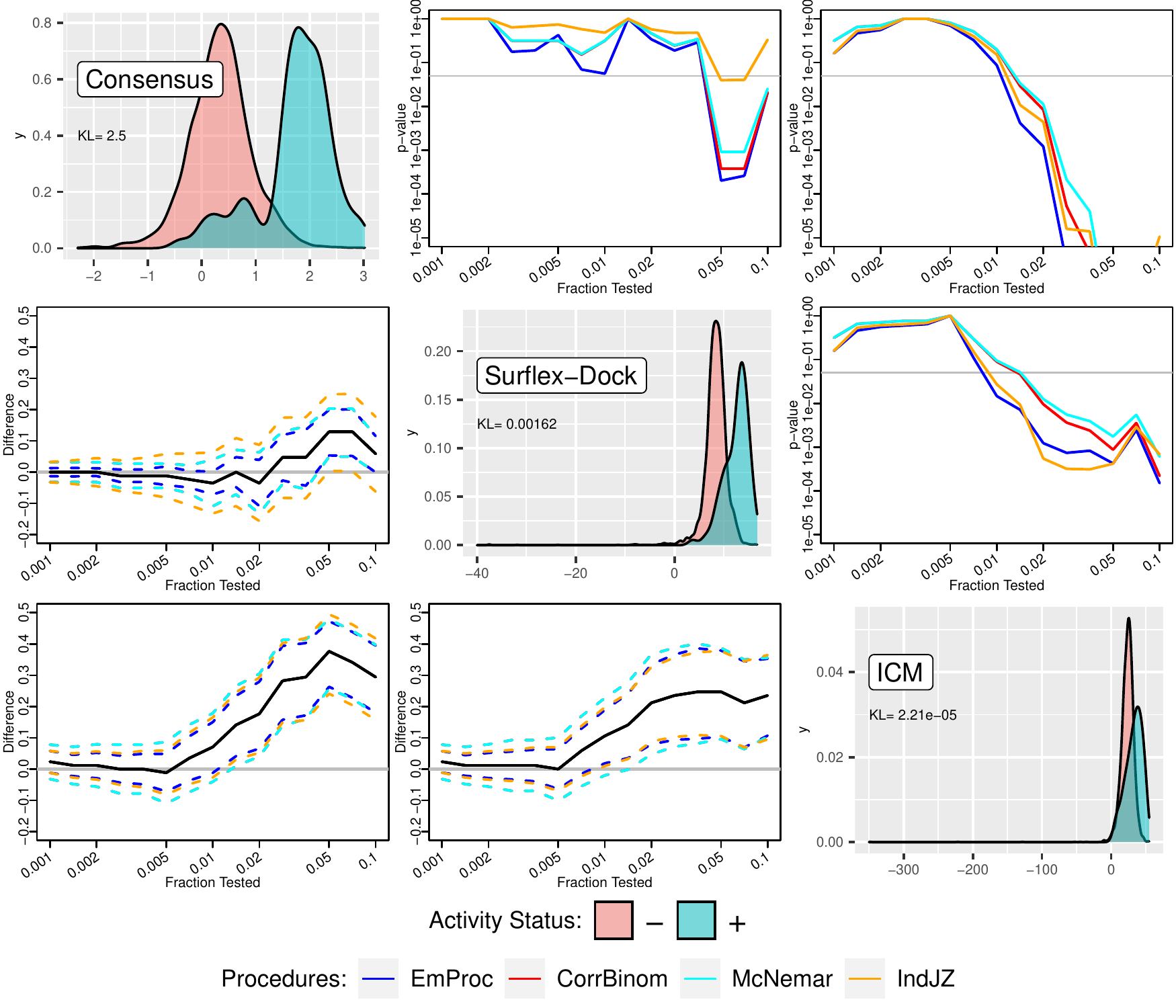}
	\caption{Comparisons of scoring methods surflex-dock, ICM, and their consensus, across 15 testing fractions using four testing procedures. The diagonal grid shows estimated score densities within activity classes for each scoring method; also shown are Kullback-Leibler divergences between estimated densities for the $+$ and $-$ classes. The upper grid shows $p$-values from testing for equality of hit enrichment curves from a pair of scoring methods, with colors corresponding to different testing procedures. The lower grid shows pointwise 95 percent confidence intervals that accompany results from the upper grid. Comparisons have not been adjusted for multiple testing as was the case in Table \ref{tab:applic}.\label{fig:panel}}
\end{figure}

Ignoring the need for multiplicity adjustments, Figure \ref{fig:panel} provides information similar to Table \ref{tab:applic}, except for many more testing fractions. The upper grid shows (unadjusted) $p$-values from testing for equality of hit enrichment curves, while the lower grid shows (unadjusted) pointwise 95 percent confidence intervals of differences between hit enrichment curves. EmProc is the only method that is consistently among the best performers. The Pearson correlation is largest (but still only moderate at 0.624) between scores from Surflex-dock and the consensus method, so the poor performance of IndJZ for this comparison is not surprising. On the other hand, the performance of IndJZ understandably improves when comparing Surflex-dock and ICM, with Pearson correlation of only 0.156 between scores. 

The diagonal grid of Figure \ref{fig:panel} shows estimated score densities within activity classes for each scoring method. The consensus method has the biggest separation of activity-class densities, with Kullback-Leibler divergence \citep{KL1951} of 2.50 compared to the substantially smaller values of 0.00162 for surflex-dock and $2.21\times 10^{-5}$ for ICM, and this is consistent with the consensus scoring method seeming to outperform the others. 

\begin{figure}[t]
	\centering\includegraphics[width=.8\textwidth]{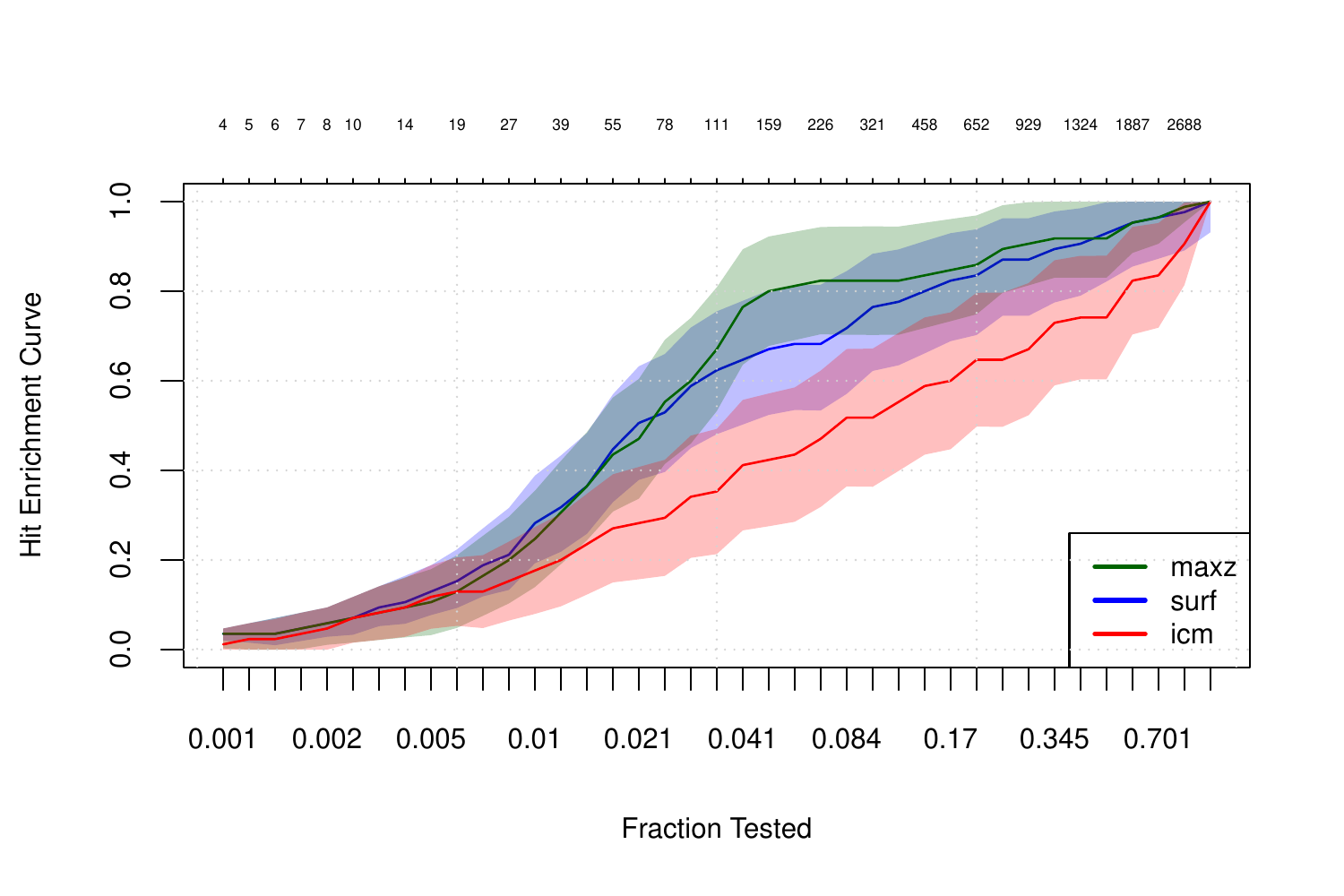}
	\caption{Simultaneous 95 percent plus-adjusted sup-t confidence bands for the three scoring methods surflex-dock, ICM, and their consensus. Even while spreading interest across the entire range of testing fractions, significant differences are still detected between the consensus and ICM for some intermediate testing fractions.\label{fig:confbands}}
\end{figure}

Taking an alternative approach, Figure \ref{fig:confbands} shows plus adjusted sup-t confidence bands for the three scoring methods. These bands account for correlation between recall values at distinct testing fractions for a single curve, and offer simultaneous coverage across the curve. While they do not account for correlations between curves, they are helpful visualizations of uncertainty that go beyond simply graphing the curves alone.

\begin{figure}
	\centering\includegraphics[width=\textwidth]{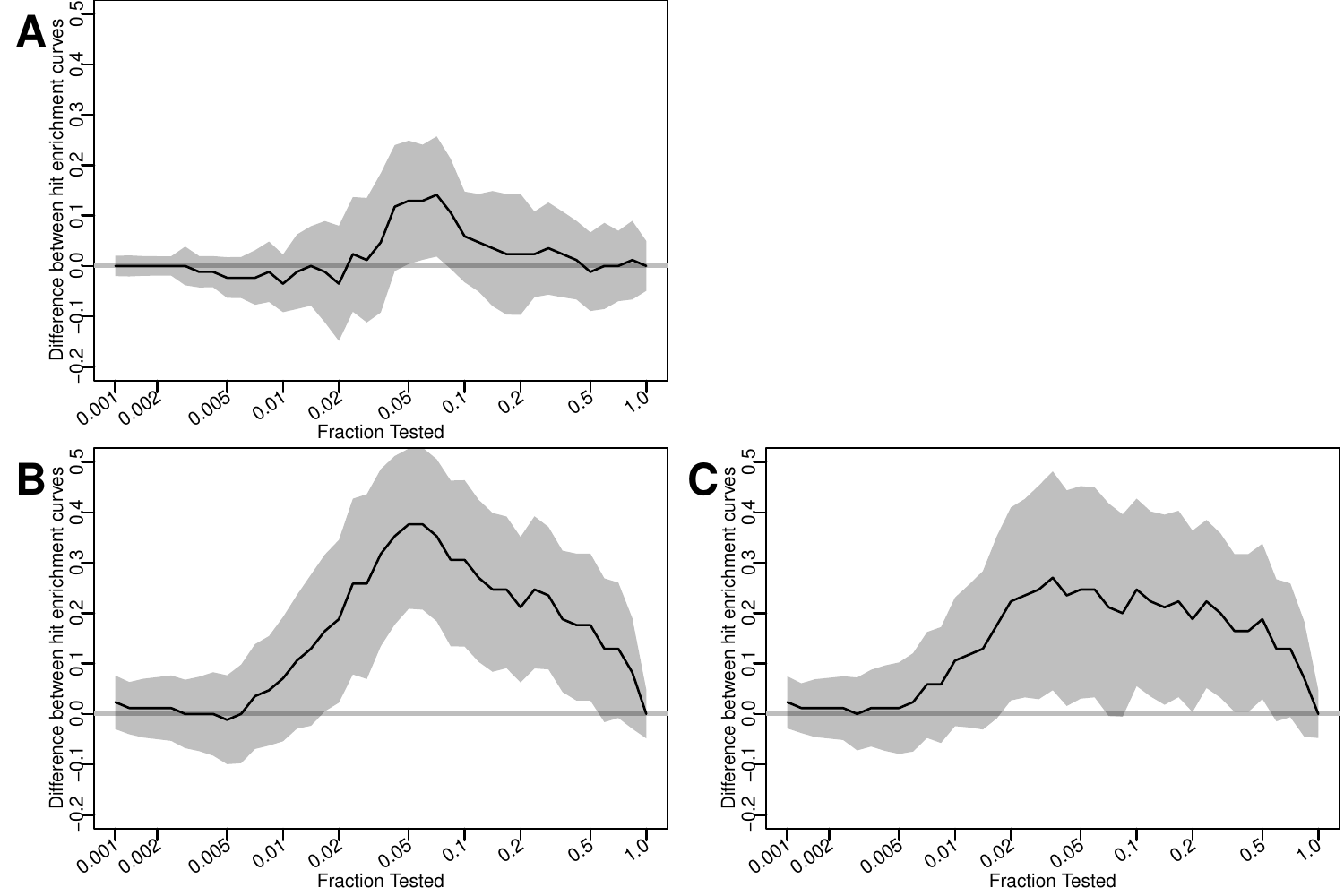}
	\caption{Simultaneous 95 percent plus-adjusted sup-t confidence bands for pairwise differences between the three scoring methods surflex-dock, ICM, and their consensus. A: consensus versus surflex-dock, B: consensus versus ICM, C: surflex-dock versus ICM. The bands are unsurprisingly wider than the pointwise EmProc intervals shown in the lower panel of Figure \ref{fig:panel}, but they are not very much wider. The ICM scoring method is significantly less effective than both consensus and suflex-dock for testing fractions between 0.02 and 0.5.\label{fig:confbands2}}
\end{figure}

A more direct approach to pairwise comparisons between curves, while adjusting for the many comparisons that occur along the curve, is shown in Figure \ref{fig:confbands2}. These bands for the differences between hit enrichment curves offer simultaneous coverage across the curve. They account for correlation between recall values at distinct testing fractions for a single curve, and for correlation between estimated recall from different scoring methods.

\section{Discussion}\label{sec:discuss}

Early enrichment is broadly recognized as a primary goal of virtual screening \citep{Truchon2007}. There is, however, considerable debate about how to assess achievement of this goal \citep{robinson2020validating}. The receiver operating characteristic (ROC) curve, and its summary metric of area under the curve, lack sensitivity to both the early recognition goal and the rarity of active compounds that often exists in screening studies \citep{Truchon2007,saito2015precision}. Partial area under the ROC curve makes a step towards addressing the early recognition goal but not the rarity of active compounds.

The robust initial enhancement (RIE, \cite{Sheridan2002}) and its normalized version the Boltzmann enhanced discrimination of ROC (BEDROC, \cite{Truchon2007}) directly address early recognition by incorporating an exponential weight that decreases for active compounds that are discovered later in testing. The versions of RIE and BEDROC that are proposed by \cite{Truchon2007} are computationally efficient. They are, however, global measures that are not tied to a specific testing fraction, and they depend on a parameter $\alpha$ that must be specified by the user. A larger $\alpha$ provides greater weight to active ligands found early, but weights are applied to all active ligands, even those found very late. Moreover, statistical inference is not straightforward. 

For the PPARg study using the default value of $\alpha=20$, the \cite{Empereur-Mot2016} web application reports BEDROC values as follows, where larger is better: 0.743 for the consensus; 0.687 for Surflex-dock; and 0.447 for ICM. Without uncertainty measures associated with these scores, it is difficult to conclude separation of the scoring methods. Furthermore, the results presented in Table~\ref{tab:applic} suggest conclusions that are more nuanced, with no significant differences at testing fraction 0.01 but some big differences at testing fraction 0.1. The improvement in BEDROC for the consensus method is mostly driven by the improvement near testing fraction 0.1, but this is a much larger fraction that what is typically used in screening evaluations.
This illustrates how BEDROC can provide a misleading sense of improvement in early enrichment. It averages performance over all testing fractions and weights early fractions more heavily, but it is often difficult to determine from the tuning parameter alone to what extent the early fractions are weighted. 

The hit enrichment curve offers multiple benefits for assessing early enrichment. First, it directly addresses the measure of interest, namely enrichment. For a given dataset, the closer the estimated hit enrichment curve is to the ideal hit enrichment curve, the more desirable is the associated algorithm. Second, the user is able to directly enforce their definition of ``early'' by specifying testing fractions of interest, without needing to rely on the indirect methods offered through measures such as BEDROC or partial area under the ROC curve. And third, localized assessment is possible, rather than whole-curve assessment. These are the reasons we have chosen to focus this work on hit enrichment curves.

Enrichment factors, and the associated enrichment factor curve, are also very popular for assessing early enrichment. As it turns out, all results in this paper can be trivially modified to obtain inference for the enrichment factor curve; see \cite{ash2020methods} for further details. While the hit enrichment curve plots $\{ r,\widehat{\theta}_r\}$, the enrichment factor curve plots $\{ r,\widehat{\theta}_r/r\}$, so standard error expressions are easily modified by division by the testing fraction $r$. Enrichment factors (sometimes denoted $EF_r$) simply focus on the enrichment factor curve at a specific testing fraction.

In this article, we provide a template for rigorously comparing competing algorithms while accounting for uncertainty, two types of correlation, and the multiple testing issue. If interest is restricted to comparing hit enrichment curves for competing algorithms at a few pre-selected testing fractions, the hypothesis testing and confidence interval procedures of Section \ref{sec:compare2} offer effective strategies. On the other hand, while it is best practice to decide testing fractions \textit{a priori}, we acknowledge this is not always done or even possible, so we also provide confidence bands to compare entire hit enrichment curves. Additionally, bands allow agmented graphical presentation of the entire hit enrichment curves for competing algorithms.

The EmProc procedure is newly proposed here and is expected to perform as well or better than the other procedures considered (CorrBinom, McNemar, and IndJZ) in the presence of correlation between competing algorithms and/or correlation across different testing fractions within a single algorithm.  The other procedures considered address only a single type of correlation
and hence are not recommended for general-purpose use. 

Finally, these inferential procedures are conveniently made available in R package \verb+chemmodlab+ on CRAN and GitHub.

\bigskip
\begin{center}
{\large\bf SUPPLEMENTARY MATERIAL}
\end{center}
\begin{description}

\item[Appendix:] Proof of Theorem 1


\end{description}

\bibliographystyle{agsm}
{\small \bibliography{HitEnrich}}
\end{document}


\begin{center}
    {\bf SUPPLEMENTARY MATERIAL}\\\ \\
    {\bf Appendix: Proof of Theorem 1}
\end{center}

Let $\textbf{W}_i = (X_i, S_{1i}, S_{2i})$, where $\textbf{W}_1, ..., \textbf{W}_n$ is a random sample from a probability distribution $P$ on a measurable space $(\mathcal{W}, \mathcal{A})$ for $\mathcal{W} = \{0, 1\} \times \mathbb{R} \times \mathbb{R}$. To arrive at the convergence in distribution result for $\widehat{\theta}_{1r} -\widehat{\theta}_{2r}$, we need to establish Hadamard
differentiability of $\widehat{\theta}_{1r} -\widehat{\theta}_{2r}$. To do
so, we use the fact that under Conditions 1 and 2, both $\widehat{\theta}_{1r}$ and
$\widehat{\theta}_{2r}$ are Hadamard differentiable \citep{Jiang2015}.
Hadamard differentiability of $\widehat{\theta}_{1r} -\widehat{\theta}_{2r}$
is implied by the definition of Hadamard differentiability (3.9.1, 
\cite{VanderVaart1996}).

The influence function associated with $\widehat{\theta}_{jr}$ is $L_j(\bm{W})=H_j$, where $H_j=(X-\Lambda_{jr})(A_{jr}-\theta_{jr})/\pi_+$ and $A_{jr}=\mbox{I}(S_j>t_{jr})$, for $j=1,2$. Applying the chain rule for influence functions, we obtain the influence function for $\widehat{\theta}_{1r} -\widehat{\theta}_{2r}$ as $L(\bm{W})=H_1-H_2$. Consequently,
\[ \sqrt{n}\Bigr\{(\widehat{\theta}_{1r} -\widehat{\theta}_{2r}) - (\theta_{1r} - \theta_{2r})\Bigl\} \xrightarrow[]{d} N(0, Var(H_1-H_2)) \]
as $n \rightarrow \infty$.
The required variance expression is obtained as $Var(H_1-H_2) = Var(H_1) + Var(H_2) -2Cov(H_1,H_2)$. Jiang \& Zhao \cite{Jiang2015} already obtained the expression for $Var(H_j)$ as
\[ \frac{\theta_{jr}(1-\theta_{jr})}{\pi_+}\Big[1 - 2\Lambda_{jr} + \frac{\Lambda^2_{jr}(1-r)r}{{\pi}_+\theta_{jr}(1-\theta_{jr})}\Big]. \]

We now obtain $Cov(H_1,H_2)$. Using the law of total covariance, we have that: 
$$
\begin{aligned}
	&Cov\big( \left(X-\Lambda_{1r}\right)\left(A_{1r} - \theta_{1r}\right), \left(X-\Lambda_{2r}\right)\left(A_{2r} - \theta_{2r}\right)\big) \\
	&= E\big[Cov\big(\left(X-\Lambda_{1r}\right)\left(A_{1r} - \theta_{1r}\right), \left(X-\Lambda_{2r}\right)\left(A_{2r} - \theta_{2r}\right) \ | \ X\big)\big] \\
	&\quad + Cov\big(E\left[\left(X-\Lambda_{1r}\right)\left(A_{1r} - \theta_{1r}\right) \ | \ X\right], E\left[\left(X-\Lambda_{2r}\right)\left(A_{2r} - \theta_{2r}\right) \ | \  X\right]\big).
\end{aligned}
$$ 

\noindent For the first term we have that: 
$$
\begin{aligned}
	&E\big[Cov\big(\left(X-\Lambda_{1r}\right)\left(A_{1r} - \theta_{1r}\right), \left(X-\Lambda_{2r}\right)\left(A_{2r} - \theta_{2r}\right) \ | \ X\big)\big] \\
	&= \pi_+ Cov\big(\left(X-\Lambda_{1r}\right)\left(A_{1r} - \theta_{1r}\right), \left(X-\Lambda_{2r}\right)\left(A_{2r} - \theta_{2r}\right) \ | \ X=1\big) \\
	&\quad + \left(1-\pi_+\right) Cov\big(\left(X-\Lambda_{1r}\right)\left(A_{1r} - \theta_{1r}\right), \left(X-\Lambda_{2r}\right)\left(A_{2r} - \theta_{2r}\right) \ | \ X=0\big) \\
	&= \pi_+ \left(1-\Lambda_{1r}\right)\left(1-\Lambda_{2r}\right)Cov\left(A_{1r}, A_{2r}|X = 1\right) 
	+ \left(1-\pi_+\right) \left(\Lambda_{1r}\right)\left(\Lambda_{2r}\right)Cov\left(A_{1r}, A_{2r}|X = 0\right) \\ 
	&= \pi_+ \left(1-\Lambda_{1r}\right)\left(1-\Lambda_{2r}\right)\left(\theta_{12\cdot r} - \theta_{1r} \theta_{2r}\right) 
	+ \left(1-\pi_+\right) \left(\Lambda_{1r}\right)\left(\Lambda_{2r}\right)\left(\theta_{12\cdot r}' - \theta_{1r}' \theta_{2r}'\right), \\
\end{aligned}
$$ 

\noindent where $\theta_{jr}' = P(S_j > t_{jr} | -)$ for
$j \in \{1, 2\}$ and $\theta_{12\cdot r}' = P(S_1 > t_{1r} , S_2 > t_{2r} | -)$. For the second term we have: 
$$
\begin{aligned}
	&Cov\big(E\left[\left(X-\Lambda_{1r}\right)\left(A_{1r} - \theta_{1r}\right) \ | \ X\right], E\left[\left(X-\Lambda_{2r}\right)\left(A_{2r} - \theta_{2r}\right) \ | \  X\right]\big) \\
	&= E\big(E\left[\left(X-\Lambda_{1r}\right)\left(A_{1r} - \theta_{1r}\right) \ | \ X\right] \cdot E\left[\left(X-\Lambda_{2r}\right)\left(A_{2r} - \theta_{2r}\right) \ | \  X\right]\big) \\
	& \quad - \left(E\left[E\left[\left(X-\Lambda_{1r}\right)\left(A_{1r} - \theta_{1r}\right) \ | \ X\right]\right]\right) \cdot \left(E\left[E\left[\left(X-\Lambda_{2r}\right)\left(A_{2r} - \theta_{2r}\right) \ | \  X\right]\right]\right) \\
	&= \left(1-\pi_+\right) \cdot E\left[\left(X-\Lambda_{1r}\right)\left(A_{1r} - \theta_{1r}\right) \ | \ X = 0\right] \cdot E\left[\left(X-\Lambda_{2r}\right)\left(A_{2r} - \theta_{2r}\right) \ | \  X= 0\right] \\
	&\quad - \left(1-\pi_+\right)E\left[\left(X-\Lambda_{1r}\right)\left(A_{1r} - \theta_{1r}\right) \ | \ X = 0\right] \cdot \left(1-\pi_+\right)E\left[\left(X-\Lambda_{2r}\right)\left(A_{2r} - \theta_{2r}\right) \ | \  X= 0\right] \\
	&= \left(\pi_+\right)\left(1-\pi_+\right)\Lambda_{1r}\Lambda_{2r}\left(\theta_{1r}' - \theta_{1r}\right)\left(\theta_{2r}' - \theta_{2r}\right).
\end{aligned}
$$ 

\noindent Thus, \begin{alignat*}{2}
	Cov(H_1,H_2) &= \frac{1}{\pi_+^2}&& \Biggl\{ \pi_+ \left(1-\Lambda_{1r}\right)\left(1-\Lambda_{2r}\right)\left(\theta_{12\cdot r} - \theta_{1r} \theta_{2r}\right) 
	+ \left(1-\pi_+\right) \left(\Lambda_{1r}\right)\left(\Lambda_{2r}\right)\left(\theta_{12\cdot r}' - \theta_{1r}' \theta_{2r}'\right) \\
	& && + \pi_+(1-\pi_+)\Lambda_{1r}\Lambda_{2r}\left(\theta_{1r}' - \theta_{1r}\right)\left(\theta_{2r}' - \theta_{2r}\right) \Biggr\}.
\end{alignat*} 

\noindent The expression in brackets can be expanded as a
bivariate polynomial function of $\Lambda_{1r}$ and $\Lambda_{2r}$:

\begin{alignat*}{2}
	&\pi_+(1-\Lambda_{1r} - &&\Lambda_{2r})(\theta_{12\cdot r} - \theta_{1r}\theta_{2r})\\
	&\quad+(\Lambda_{1r}\Lambda_{2r})&&\Biggl\{\pi_+(\theta_{12\cdot r} - \theta_{1r}\theta_{2r})
	+ (1-\pi_+)(\theta_{12\cdot r}' - \theta_{1r}'\theta_{2r}') \\
	& && + \pi_+(1-\pi_+)(\theta_{1r}' - \theta_{1r})(\theta_{2r}' - \theta_{2r})\Biggr\}.
\end{alignat*} 

\noindent The coefficient of $\Lambda_{1r}\Lambda_{2r}$ is
complicated but noticing that: 
$$
E[Cov(A_{1r}, A_{2r}|X)] = \pi_+(\theta_{12\cdot r} - \theta_{1r}\theta_{2r}) + (1-\pi_+)(\theta_{12\cdot r}' - \theta_{1r}'\theta_{2r}')
$$ 

\noindent and that: 
$$
\begin{aligned}
	&Cov\left(E\left[A_{1r}|X\right], E\left[A_{2r}|X\right]\right) \\
	& = E\left[E\left[A_{1r}|X\right] \cdot E\left[A_{2r}|X\right]\right] - E\left[E\left[A_{1r}|X\right]\right] \cdot E\left[E\left[A_{2r}|X\right]\right] \\
	& = \pi_+\theta_{1r}\theta_{2r} + \left(1-\pi_+\right)\theta_{1r}'\theta_{2r}' 
	- \left(\pi_+\theta_{1r} + \left(1-\pi_+\right)\theta_{1r}'\right)\left(\pi_+\theta_{2r} + \left(1-\pi_+\right)\theta_{2r}'\right)\\
	& = \pi_+\left(1-\pi_+\right)\left(\theta_{1r}' - \theta_{1r}\right)\left(\theta_{2r}' - \theta_{2r}\right)
\end{aligned}
$$ 

\noindent implies that the coefficient of $\Lambda_{1r}\Lambda_{2r}$ is
$E\left[Cov\left(A_{1r}, A_{2r}|X\right)\right] + Cov\left(E\left[A_{1r}|X\right], E\left[A_{2r}|X\right]\right) = Cov(A_{1r}, A_{2r}) = \gamma_{12\cdot r} - r^2$.

Thus: 
$$
\begin{aligned}
	& Cov\left(H_1,H_2\right) \\
	&\quad = \frac{1}{\pi_+^2}\left(\pi_+(\theta_{12\cdot r} - \theta_{1r}\theta_{2r})(1-\Lambda_{1r} - \Lambda_{2r}) + \left(\gamma_{12\cdot r} - r^2\right)\Lambda_{1r}\Lambda_{2r} \right) \\ 
	&\quad = \frac{\left(\theta_{12\cdot r} - \theta_{1r}\theta_{2r}\right)}{\pi_+} \left\{\left(1-\Lambda_{1r} - \Lambda_{2r}\right) + \frac{\left(\gamma_{12\cdot r} - r^2\right)\Lambda_{1r}\Lambda_{2r}}{\pi_+\left(\theta_{12\cdot r} - \theta_{1r}\theta_{2r}\right)}\right\}.\\
\end{aligned}
$$

\bibliographystyle{agsm}
\bibliography{HitEnrich}